\documentclass[prl,twocolumn,twoside,showpacs,superscriptaddress]{revtex4}
\usepackage{graphics,amssymb,amsmath,epsf}
\epsfclipon
\usepackage{xspace}

\def\lsim{\raise0.3ex\hbox{\,$<$\kern-0.75em\raise-1.1ex\hbox{$\sim$}\,}}
\def\gsim{\raise0.3ex\hbox{\,$>$\kern-0.75em\raise-1.1ex\hbox{$\sim$}\,}}

\begin{document}


\title{
\vspace*{-1.5cm}
\begin{flushright}
{\rm\small
SLAC-PUB-14212}
\end{flushright}
Electron in a transverse harmonic cavity}

\author{ H.~Honkanen, P.~Maris, J.~P.~Vary}
\email{heli, pmaris, jvary@iastate.edu}
\affiliation{
Department of Physics and Astronomy, Iowa State University, Ames, Iowa 50011, USA}
\author{S.~J.~Brodsky}
\email{sjbth@slac.stanford.edu}
\affiliation{
SLAC National Accelerator Laboratory, Stanford University, Menlo Park, California, USA }

\begin{abstract}
Recent experiments with heavy ions and planned experiments with ultra intense lasers 
require nonperturbative solutions to quantum field theory for predicting and 
interpreting the results. To propel this theoretical direction, we solve the 
nonperturbative problem of an electron in a strong
transverse confining potential using Hamiltonian light-front quantum field theory.
We evaluate both the invariant mass 
spectra and the anomalous magnetic moment of the lowest state for this
two-scale system. The weak external field limit of the anomalous magnetic 
moment agrees with the result of QED perturbation theory  within the 
anticipated accuracy. 
\end{abstract}
\maketitle 

Recent intense interest in strong-field dynamics,
ranging from the anomalous enhancement of lepton production in 
ultrarelativistic collisions between heavy nuclei
at RHIC \cite{Adare:2009qk} and predicted photon yield depletion at the
LHC \cite{Tuchin:2010gx},
to proposals for producing supercritical fields with
next-generation laser facilities \cite{Ruf:2009zz,Dumlu:2010ua},
points to the importance of developing new methods for solving QED  in its
nonperturbative domain.
An ideal tool for such problems is Hamiltonian light-front formalism 
[see e.g. Ref.\cite{Brodsky:1997de}], in which the gauge theory is quantized
on the light-front  and the physical states are  
expanded in a Fock-space basis developed from the constituents.
The Hamiltonian is represented as an operator acting on
these Fock states. Since time is set along the light-front, 
the ground state of the free theory is also a ground state of the full 
interacting theory
and the formalism is Lorentz frame independent. 

The light-front Hamiltonian approach
provides a realization of Feynman's covariant "parton" model where the partons
are the elementary fields used to define the Fock-space basis. 
It also has the appearance of a standard
quantum many-body problem with the necessary quantum field theory features
such as pair creation and annihilation.
Diagonalization of the Hamiltonian provides amplitudes for 
evaluating experimental observables
that are nonperturbative and relativistically invariant such as 
masses, form factors, and structure functions. The Fock-space dependence of 
observables is to be evaluated and one seeks to eliminate such dependence.

We address the problem of an electron in a transverse harmonic cavity and solve
for its mass spectra and other observables as a function of the external field 
strength over a range spanning the electron mass.
To accomplish this, we evaluate the QED Hamiltonian in light-front gauge
on the light-front in a 
Fock-space consisting of electron states and  electron plus photon states.  
 We add the harmonic 
oscillator potential in the transverse direction to confine the system in those
directions.
We then solve for the eigenvalues, eigenvectors, and anomalous magnetic moment.
The nonperturbative analysis presented in this Letter could be applicable
to measurements of the (gyromagnetic) ratio of the spin precession to Larmor 
frequencies of a trapped electron in strong external electromagnetic fields or 
intense time-dependent laser fields.
This research also serves 
as a foundation for solving other quantum field theories at strong coupling, 
such as the light-front QCD Hamiltonian in the nonperturbative domain.

The question arises on how to implement a consistent renormalization program.
Our specific choice is defined below.  With our limits on the Fock-space
and adopted renormalization scheme, we can already demonstrate the 
effects of mass renormalization but not yet coupling constant renormalization.
A full renormalization program will ensue when we
enlarge our Fock-space basis to include electron-positron pairs.
Such pairs produced at RHIC experience
strong time-dependent EM fields.  The total charge $Z_{\rm total} = Z_1 + Z_2$
of
the two colliding nuclei can exceed 137, indicating
the need to treat the production and propagation
properties with nonperturbative methods.
In particular, the renormalization scale entering the running QED coupling is
set by the photon virtuality which would be expected to be of order $ Z_{\rm
total} \alpha  \bar m_e$.
The renormalization scale $p^2$ appearing in the electron running mass
$\bar m(p^2)$ is of similar order. 
Such a full renormalization program will also allow comparison with higher 
order 
perturbative calculations such as those for an electron in a 
Penning trap \cite{Brown:1985rh}.

We define our 
light-front coordinates as $x^{\pm}=x^0 \pm x^3$, $x^{\perp}=(x^1,x^2)$,
where the variable $x^+$ is light-front time and $x^-$ is the longitudinal 
coordinate. We adopt $x^+=0$, the ``null plane", for our quantization surface.
Here we adopt 
basis states for each constituent that consist of transverse
2D harmonic oscillator (HO) states combined with discretized 
longitudinal  modes, plane waves, satisfying selected boundary conditions. 
This basis function approach follows \cite{Vary:2009gt} and is supported by 
successful 
anti-de Sitter-QCD models \cite{deTeramond:2008ht}.

The HO states are
characterized by a principal quantum 
number $n$,  orbital quantum number $m$, and HO energy 
$ \Omega $. Working in momentum space, it is convenient to write the 
2D oscillator as a 
function of the dimensionless variable  
$\rho=\vert p^\perp\vert/\sqrt{M_0\Omega}$, and 
$M_0$ has units of mass.
The orthonormalized HO wave functions in polar coordinates 
$(\rho,\varphi)$ are then given in terms of the 
generalized Laguerre polynomials, $L_n^{|m|}(\rho^2)$, by
\begin{eqnarray}
&&\Phi_{nm}(\rho,\varphi)= \langle \rho \varphi | n m \rangle\nonumber \\
&&=
\sqrt{\frac{2\pi}{M_0\Omega}}\sqrt{\frac{2n!}{(\vert m\vert+n)!}} 
e^{im\varphi}
\rho^{\vert m\vert} e^{-\rho^2/2}L^{\vert m\vert}_n(\rho^2),
\label{Eq:wfn2dHOchix}
\end{eqnarray}
with eigenvalues $E_{n,m}=(2n+|m|+1)\Omega$.
The HO wavefunctions have the 
same analytic structure in both coordinate and momentum space, a feature 
reminiscent of a plane-wave basis.

The longitudinal modes, $\psi_{k}$, in our basis are  defined for
$-L \le x^- \le L$ with  periodic boundary conditions for the photon and 
antiperiodic boundary conditions for the electron:
\begin{eqnarray}
  \psi_{k}(x^-) &=& \frac{1}{\sqrt{2L}} \, {\rm e}^{i\,\frac{\pi}{L}k\,x^-},
\label{Eq:longitudinal1}
\end{eqnarray}
where $k=1,2,3,...$  for periodic boundary conditions
 (we neglect the zero mode) and 
$k=\frac{1}{2},\frac{3}{2},\frac{5}{2},...$  for
antiperiodic boundary conditions. 
The full 3D single-particle basis state is defined by the product form
\begin{eqnarray}
  \Psi_{k,n,m}(x^-,\rho,\varphi) &=& \psi_{k}(x^-) \Phi_{n,m}(\rho,\varphi).
\label{Eq:totalspwfn}
\end{eqnarray}

Following Ref.\cite{Brodsky:1997de} we introduce the total invariant
 mass-squared $M^2$ for 
the low-lying physical states in terms of a Hamiltonian $H$ times a 
dimensionless integer for the total light-front momentum $K$
\begin{eqnarray}
M^2 + P_{\perp}P_{\perp} \rightarrow  M^2 + const = P^+P^- = KH
\label{Mass-squared}
\end{eqnarray}
where we absorb the constant into $M^2$.  
For simplicity, the transverse functions for both the electron and the photon 
are taken as eigenmodes of the trap.
The noninteracting Hamiltonian $H_0= 2M_0 P^-_c$ for this system is then defined by the 
sum of the occupied 
modes $i$ in each many-parton state:
\begin{eqnarray}
&& H_0 = \frac{2M_0\Omega}{K}\sum_i{\frac{2n_i+|m_i| +1 +
{{\bar m_i}^2}/(2M_0\Omega)
}{x_i}},
\label{Hamiltonian}
\end{eqnarray}
where ${\bar m_i}$ is the mass of the parton $i$. 
The photon mass is set to zero throughout this work and the electron
mass $\bar m_e$ is set at the physical mass 0.511 MeV in our 
nonrenormalized 
calculations. We also set  $M_0=\bar m_e$.

The light-front QED Hamiltonian interaction terms we need are
the electron to electron-photon vertex, given as
\begin{equation}
V_{e\to e\gamma}=g\int dx_{+}d^2x_{\perp}
\overline{{\Psi}}(x)\gamma^\mu {\Psi}(x){A}_\mu(x)
\bigg\vert_{x^{+}=0},
\label{vertex}
\end{equation}
and the instantaneous electron-photon interaction,
\begin{equation}
V_{e\gamma\to e\gamma}=\frac{g^2}{2}\int\!dx_+d^2x_{\!\perp}
    \ \overline{\Psi}\gamma^\mu 
    A_\mu \ \frac{\gamma^+}{i\partial^+}
    \left( \gamma^\nu  A_\nu 
    \Psi  \right)\bigg\vert_{x^{+}=0},
\label{inst}
\end{equation}
where the coupling constant $g^2=4\pi\alpha$, and $\alpha$ is the fine 
structure constant taken to be $\alpha=\frac{1}{137.036}$ in this work.
The nonspinflip
vertex terms of Eq.(\ref{vertex}) are  $\propto M_0\Omega $, whereas
spinflip terms are $\propto \sqrt{M_0\Omega} m_e$. 
Selecting the initial state electron helicity in the single electron sector
always as ``up'' the process $e\to e\gamma$ is nonzero for three out of eight
helicity combinations, and the process 
 $e\gamma\to e\gamma$ is nonzero only with all four spin projections aligned 
(two out of 16 combinations), resulting in a sparse matrix.

We implement a symmetry constraint for the basis by fixing the total angular 
momentum projection 
$J_z=M+ S=\frac{1}{2}$,
where  $M=\sum_i{m_i}$ is the total azimuthal quantum number, and  
$S=\sum_i{s_i}$
the total spin projection along the $x^-$ direction.  
For cutoffs, we select the total light-front momentum, $K$, and the maximum 
total quanta allowed in the transverse mode of each one or two-parton state, 
$N_{max}$, such that 
\begin{eqnarray}
&&\sum_i{x_i} = 1= \frac{1}{K}\sum_i{k_i},\label{momsumrule} \\
&&\sum_i{2n_i+|m_i| +1} \le N_{max},
\end{eqnarray}
where, for example, $k_i$ defines the longitudinal 
modes of Eq.(\ref{Eq:longitudinal1}) for the $i^{th}$ parton. 
Equation (\ref{momsumrule}) signifies total light-front momentum conservation 
written in terms of boost-invariant momentum fractions, $x_i$.
Since we 
employ a mix of boundary
conditions and all states have half-integer 
total K, we will quote  K values rounded downwards for convenience, except 
when the precise value is required.

\begin{figure}
\epsfxsize=9.cm
\epsfbox{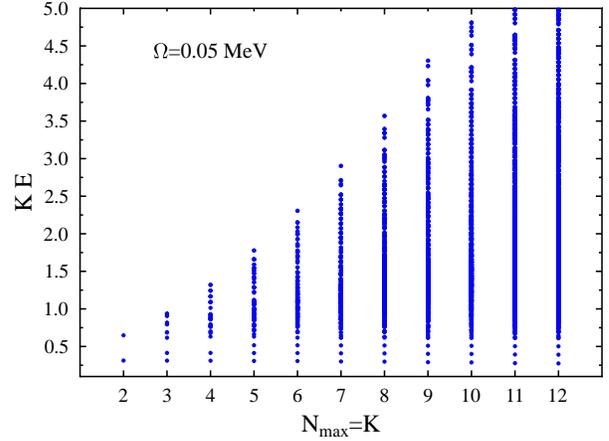}
\caption{\label{fig1} (color online). Eigenvalues (multiplied by $K$) for a 
nonrenormalized
light-front QED Hamiltonian which includes the electron-photon vertex
and the instantaneous electron-photon interaction without counterterms. 
The cutoffs
for the basis space dimensions are selected such that $K$ 
increases simultaneously with the $N_{max}$.}
\end{figure}

In Fig.\ref{fig1} we show the eigenvalues (multiplied by $K$) for a 
nonrenormalized light-front QED Hamiltonian given in 
Eqs.(\ref{Hamiltonian},\ref{vertex},\ref{inst}), with fixed $\Omega=0.05$ MeV
and simultaneously increasing $K$ and $N_{max}$.
The resulting dimension of the Hamiltonian matrix increases rapidly. 
For $N_{max}=K=2,10,$ and $20$,
the dimensions of the corresponding symmetric $d\times d$ matrices are 
$d=2,\, 1670,$ and $26\, \,990$, respectively.

The number of the single electron basis states, considering all the symmetries,
 increases slowly with increasing 
$N_{max}=K$ cutoff. For  $N_{max}=K=2,10,$ and $20$ the number of single electron basis
states is $1,5,$ and $10$, respectively.
Our lowest-lying eigenvalue corresponds to a solution dominated by the electron
with $n=m=0$. The ordering of excited states, due to significant interaction 
mixing, does not always follow the highly degenerate unperturbed spectrum of 
Eq.(\ref{Hamiltonian}).  States dominated by spin-flipped electron-photon  
components are evident in the solutions.
Nevertheless, the  lowest-lying eigenvalues appear with nearly harmonic 
separations in  Fig.\ref{fig1} as would be expected at the coupling of QED.
The multiplicity of the higher eigenstates increases rapidly with increasing
$N_{max}=K$ and the states exhibit stronger mixing with other states than
the lowest-lying states. In principle, the electron-photon basis 
states interact directly with each other in leading order through the 
instantaneous
electron-photon interaction, but numerically the effect of this interaction
is very weak, and thus does not contribute significantly to the mixing.
Even though we work within a
Fock-space approach, our numerical results should 
approximate the lowest
order perturbative QED results for sufficiently weak external field.
\begin{figure}
\epsfxsize=9.cm 
\epsfbox{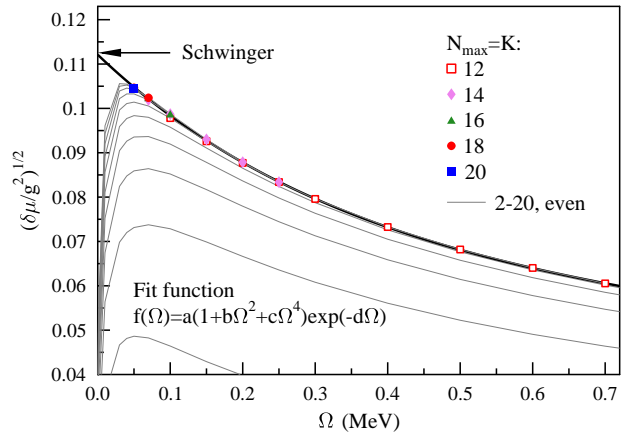} 
\caption{\label{fig2} (color online). 
Square root of the (scaled) electron anomalous magnetic 
moment as a function of the transverse external field for a sequence of 
increasing basis spaces (solid grey lines). 
These are nonrenormalized results 
where the mass eigenvalue falls within 25\% of the free electron mass.
The theoretical one-loop QED 
prediction (``Schwinger'') result  of 0.1125, appropriate to $\Omega=0$ MeV,
is indicated.
The black solid line is a fit to the
results for $N_{max}=K=12,\dots 20$, points included into the fit are indicated
by the markers in the legend.
Extrapolation to zero external field yields 0.1121.}
\end{figure}

In Fig.\ref{fig2} we show the results for the square root of the electron 
anomalous magnetic moment (scaled),  
$\sqrt{\delta\mu / g^2}$,
as a function of $\Omega$ obtained from the
lowest mass eigenstate. 
That is, we plot the magnitude of the probability 
amplitude that electron has its spin flipped relative to the single electron 
Fock-space component in the range where the results are converged.
Since our system is in an external field, the lowest physical mass 
eigenstate (not known experimentally) can deviate from the free electron mass. 
Therefore, without renormalization, we only consider cases
where the mass eigenvalue falls within 25\% of the free electron mass.
At zero external field we may compare our $\delta\mu$ to
the QED one-loop contribution to the electron anomalous magnetic moment,
the Schwinger result $\frac{\alpha}{2\pi}$ \cite{Schwinger:1948iu}. That is
we compare our results with 
$\sqrt{{\alpha}/{(2\pi g^2)}}= \sqrt{{1}/{8\pi^2}}$.
For even $N_{max}=K$ the results converge 
rapidly for  $N_{max}=K\ge 14$. The results for odd cutoffs (not shown) track
even cutoff results as $N_{max}=K$ increases. Below $\Omega\lsim 0.05$ MeV
all interactions are quenched 
at fixed $N_{max}=K$, and not converged,  due in part to  our 
requirement that the HO basis tracks the external field.

Figure \ref{fig2} also shows an extrapolation of the above
results  for $N_{max}=K=12,\dots,20$  to the
zero external field limit $\Omega=0$ Mev. We have only included the points
 above the peak at $\Omega\gsim 0.05$ MeV,
where we have reasonable convergence.
An excellent agreement with the 
results is obtained by a fit function $f(\Omega)=
a(1+b\Omega^2+c\Omega^4)\exp(-d\Omega)$, with $a=0.1121$. This is 
$<1\%$ deviation from the Schwinger 
result of 0.1125, which is reasonable in light of our numerical accuracy and extrapolation uncertainties.
If we perform individual extrapolations for all the
$N_{max}=K=12,\dots,20$ results with $0.1\le \Omega\le 1.4$ MeV, 
a range spanning the electron mass scale, we obtain 
excellent 
fits with $0.1109 \le a \le 0.1134$, i.e., remaining within  1.5\%  of the
Schwinger result.

In Ref.\cite{Vary:2009gt} we discussed  possible divergences present
in our framework,
 and anticipated a straightforward  management
of the identified divergences. 
Here we renormalize our results
by applying a sector-dependent normalization scheme from 
Ref.\cite{Karmanov:2008br}.
In our present limited Fock-space, we need only the mass counterterm 
$\delta m_e$. This $\delta m_e$ is added to the mass term in the
diagonal one-electron part of 
the Hamiltonian Eq.(\ref{Hamiltonian}).
In the absence of a known  experimental mass for
renormalization due to the external field, we adjust 
$\delta m_e$ such that the lowest 
eigenstate remains at $K E_0={m_e^2+M_0\Omega}$. That is, we simply 
adopt the free electron mass for the renormalized mass, and keep the
coupling constant $g^2$ unchanged. We emphasize that our choice for the 
renormalized mass and for the coupling constant are, in principle, valid
for the case of zero external field only. Measurements for electrons
in a trap (see, e.g., \cite{Hanneke:2010au})
 could provide results leading to more
precise renormalization parameters, but this aspect is beyond the scope of
this Letter.
\begin{figure}
\epsfxsize=9.cm
\epsfbox{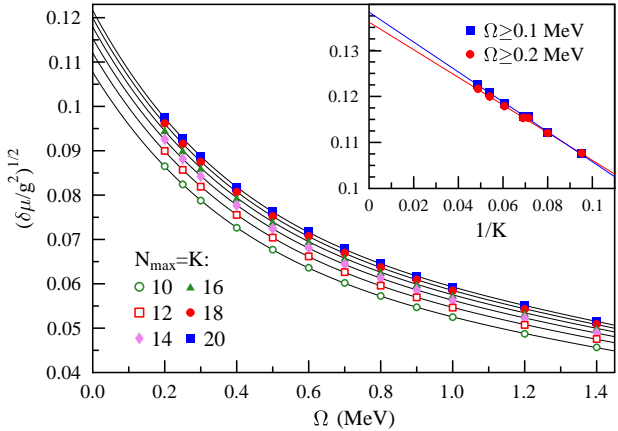}
\caption{\label{fig4} (color online). Individual  fits (solid black lines)
to the  renormalized 
results for square root of the (scaled) electron anomalous magnetic 
moment for $N_{max}=K=10,\dots,20$.
 The inset shows
the continuum limit extrapolation of the zero external field 
results in the main panel as a function of $1/K$.}
\end{figure}

In Fig.\ref{fig4} we present $\sqrt{\delta\mu / g^2}$
for $N_{max}=K=10,\dots,20$ from the 
renormalized QED Hamiltonian of
Eq.(\ref{Hamiltonian}), with $\delta m_e$, and 
Eqs.(\ref{vertex},\ref{inst}). To eliminate possible effects 
from the peak at $\Omega\sim 0.05$ MeV in Fig.\ref{fig2}, we only include
results with the external field $\Omega\ge 0.2$ MeV. Again, individual
fits of the form  $f(\Omega)=a(1+b\Omega^2+c\Omega^4)\exp(-d\Omega)$ are 
an excellent representation of our results. The range of the extrapolated 
values is $0.1077 \le a \le 0.1216$.

The convergence with an 
increasing cutoff is now less rapid than in the nonrenormalized case shown
in Fig.\ref{fig2}. In order to approach the continuum limit 
$N_{max}=K\to\infty$,
we perform further extrapolation to the zero-$\Omega$  results
of Fig.\ref{fig4}.
The inset of Fig.\ref{fig4} shows
linear extrapolation of the results of the main 
figure in $1/K$ to the continuum
limit  $N_{max}=K\to\infty$. To verify the stability of the results, an 
extrapolation based on the  $\Omega\ge 0.1$ MeV fits (not shown)
is also given.
The extrapolated continuum values are 0.1362 (0.1383) for  
$\Omega\ge 0.2\, (0.1)$, respectively, and thus about 20\%
above the Schwinger result 0.1125. An enhancement of this 
magnitude was also observed in related works, 
Refs.\cite{Brodsky:1998hs,Brodsky:2004cx}  and 
Refs.\cite{Chabysheva:2009ez,Chabysheva:2009vm}, where the one-photon truncated
light-front Hamiltonian was regulated with Pauli-Villars 
regularization scheme. With  Pauli-Villars regularization as well as in our 
renormalized 
results, interpreted from a perturbation theory perspective, the intermediate 
state propagators are developed from a dynamical (nonperturbative) electron 
mass rather than using the unperturbed mass needed for direct comparison with 
perturbation theory.

In our approach, the HO parameters $\Omega,M_0$, the electron mass $m_e$, and 
the total longitudinal momentum $K$ appear as prefactors for the matrix
elements in the Hamiltonian. Therefore, we can rather 
straightforwardly vary the size of 
the Hamiltonian matrix by keeping  $N_{max}$ fixed, and changing $K$ alone. 
We studied the continuum limit of  
$\sqrt{\delta\mu / g^2}$ by
setting  $N_{max}=20$ and increasing $K$ in units of 10, from $K=10$ to $K=50$.
The dimension of the Hamiltonian matrix then increases from $d=11790$ to 
$d=69590$. The extrapolated results range between 
$0.1148 \le a \le 0.1259$, and show a good convergence pattern.
Linear extrapolation of these results, analogously to Fig.\ref{fig4},  
to the continuum limit  $K\to\infty$
are  0.1288 (0.1290) 
with $\Omega\ge 0.2\, (0.1)$ MeV, $\sim 15\%$ above the Schwinger
value.

In summary, we have evaluated properties of an electron in a nonperturbative 
external harmonic oscillator potential.  We have taken the
weak external field limit of the electron
anomalous magnetic moment, and obtained results compatible with QED 
perturbation theory  with reasonable accuracy.
Our framework can be extended by 
incorporating higher Fock-space sectors and adopting external strong fields 
relevant to heavy ion collisions and to future high-intensity laser facilities.  Applications to QCD will 
proceed with the adoption of recently developed color-singlet basis 
enumeration techniques \cite{Vary:2009gt}.

The authors thank A. Harindranath, K.~Tuchin, J.~Hiller, S.~Chabysheva, 
V.~Karmanov and A.~Ilderton for fruitful discussions.  
Computational resources were provided by DOE through the
National Energy Research Supercomputer Center (NERSC).
This work was supported in part by a DOE Grant No. DE-FG02-87ER40371 and by DOE Contract No. DE-AC02-76SF00515.


\end{document}